# Grafted AlGaAs/GeSn Optical Pumping Laser Operating up to 130 K

Jie Zhou,[1,†] Daniel Vincent,[1,†] Sudip Acharya,[2,3,†] Solomon Ojo,[2,3,†] Alireza Abrand,[4] Yang Liu,[1] Jiarui Gong,[1] Dong Liu[1], Samuel Haessly,[1] Jianping Shen,[1] Shining Xu,[1] Yiran Li,[1] Yi Lu,[1] Hryhorii Stanchu,[2] Luke Mawst,[1] Bruce Claflin,[5] Parsian K. Mohseni,[4,*] Zhenqiang Ma[1,*] and Shui-Qing Yu[2,*]

[1]Department of Electrical and Computer Engineering, University of Wisconsin - Madison, Madison, 53706, USA
[2]Department of Electrical Engineering, University of Arkansas, Fayetteville, Arkansas 72701, USA
[3]Microelectronics-Photonics Program, University of Arkansas, Fayetteville, Arkansas 72701, USA
[4]Department of Electrical and Microelectronic Engineering, Rochester Institute of Technology, Rochester, New York 14623, USA
[5]Air Force Research Laboratory, Wright-Patterson AFB, OH 45433, USA
[†]The authors contributed equally to this letter
*mazq@engr.wisc.edu
*syu@uark.edu
*pkmeen@rit.edu

**Group IV GeSn double-heterostructure (DHS) lasers offer unique advantages of a direct bandgap and CMOS compatibility. However, further improvements in laser performance have been bottlenecked by limited junction properties of GeSn through conventional epitaxy and wafer bonding. This work leverages semiconductor grafting to synthesize and characterize optically pumped ridge edge-emitting lasers (EELs) with an AlGaAs nanomembrane (NM) transfer-printed onto an epitaxially grown GeSn substrate, interfaced by an ultrathin $Al_2O_3$ layer. The grafted AlGaAs/GeSn DHS lasers show a lasing threshold of 11.06 mW at 77 K and a maximum lasing temperature of 130 K. These results highlight the potential of the grafting technique for enhancing charge carrier and optical field confinements, paving the way for room-temperature electrically injected GeSn lasers.**

**Introduction.** On-chip integrated optoelectronics utilizing group IV semiconductor materials have been a long-pursuing goal due to their low costs and ease of manufacturing considering the compatibility with existing Si-based photonics and CMOS infrastructures [1]. However, these semiconductors have faced significant challenges, particularly in light generation applications, due to the indirect bandgap nature inherent to their crystal structures. This inefficiency poses a major obstacle for the integration of group IV light emitters, such as lasers, into on-chip systems [2]. To circumvent this material challenge, research has focused on the heterogeneous integration of active layers, such as III-V semiconductors with direct bandgaps, onto mechanically supportive substrates like silicon. Techniques such as heteroepitaxy and wafer bonding/fusion have been employed to achieve this integration [2–4]. Over the years, substantial progress has been made in these approaches, and they continue to hold promise for future advancements in the field.

GeSn offers another solution to these challenges by enabling indirect-to-direct bandgap transitions while seamlessly integrating with existing silicon-based devices and CMOS fabrication processes [5–7]. As a result, GeSn has been widely investigated as an ideal candidate material for group-IV lasers [8–13]. In this field, epitaxial growth and rigid wafer bonding are the two mainstream technical approaches to realizing high-performance lasers.

In epitaxial techniques, double-heterostructures (DHS) utilizing Ge/GeSn/SiGeSn heterojunctions have been vastly employed to improve laser performance. For instance, Zhou *et al.* [9,10] demonstrated that growing a cladding layer of SiGeSn lattice-matched to GeSn with improved offset in conduction band energies produced lasers with high maximum operation temperatures and low thresholds. Additionally, multi-quantum well structures have been explored to enhance charge carrier confinement, reduce Auger recombination, and achieve population inversion in the active materials [12,14]. In the context of wafer bonding, GeSn epilayers are often bonded to dielectrics to form the GeSn-on-insulator (GeSnOI) platform, where GeSn disk lasers are constructed [13]. These wafer-bonded heterostructures achieve higher optical field confinement by leveraging the larger refractive index differences between GeSn and the buried oxides. Collectively, these studies underscore the importance of enhancing GeSn laser performance through improved charge carrier confinement and optimized light field distributions.

However, these conventional methods encounter several intrinsic challenges towards higher performance GeSn lasers. (i) Limited junction and material electronic properties. Epitaxial Ge/GeSn/SiGeSn heterojunctions exhibit modest conduction and valence band offsets as constrained by lattice matching, which restricts the potential for enhancing carrier confinement through epitaxy. Moreover, as the tin content in the GeSn alloy increases, growth quality decreases due to tin segregation in GeSn layers. (ii) Elevated density of interfacial states. Significant differences in tin content between adjacent epilayers often lead to severe lattice mismatches, resulting in a degraded interface. This, in turn, compromises charge carrier transport and reduces internal quantum efficiency. (iii) Minimal refractive index difference. The small refractive index difference between Ge/GeSn/SiGeSn heterojunctions limits optical confinement, which is crucial for efficient light-matter interaction. (iv) The increased technical complexities and manufacturing costs are concerning when creating lasers on GeSnOI platforms. Given these challenges, incorporating materials with more favorable band alignment onto epitaxially grown GeSn substrates emerges as a promising step towards achieving ideal DHS designs, ultimately paving the way for room-temperature electrically pumped GeSn lasers.

Semiconductor grafting [15–22] has been proposed as a viable approach to creating high-quality lattice-mismatched heterojunctions, overcoming many of the challenges associated with epitaxy and wafer bonding, particularly when dealing with materials that have significant discrepancies in lattice parameters and thermal properties. The grafting method enables the creation of an abrupt transition at the heterointerface of dissimilar materials while maintaining a well-suppressed density of states. This approach allows for the construction of devices that combine the favorable mechanical, optical, and electronic properties of distinct materials, thus enhancing overall device performance [16,21,22].

In this Letter, we employ this grafting method to synergistically engineer the electronic band alignment and optical field distribution by creating an AlGaAs/GeSn heterojunction. Using this heterojunction, we fabricated and characterized optically pumped lasers with ridge cavities comprising an AlGaAs nanomembrane (NM) transfer-printed onto an epitaxially grown GeSn substrate, interfaced by an ultrathin layer of $Al_2O_3$. The grafted AlGaAs/GeSn heterojunction laser features a lasing wavelength of ~2244 nm at the near-infrared range, with a lasing threshold of 11.06 mW at 77 K and a maximum lasing temperature of 130 K. This work serves as an attempt to tackle the issues related to limited junction properties in GeSn lasers and lays the foundation for future electrically injected GeSn lasers capable of operating at room temperature.

**Double-heterostructure design.** The as-grown GeSn epi structure contains a 220-nm $p^+$-$Si_{0.03}Ge_{0.89}Sn_{0.08}$ cap, a 520-nm unintentionally doped (UID) $i$-$Ge_{0.90}Sn_{0.10}$ active layer, a 650-nm $n^+$-$Ge_{0.93}Sn_{0.07}$ buffer layer, and a 430-nm $n^+$-Ge buffer layer, all epitaxially grown on a Si bulk substrate using a chemical viper deposition (CVD) chamber. Secondary Ion Mass Spectrometry (SIMS) was first conducted to determine the doping and elementary profile of the as-grown GeSn substrate, as seen in Fig. 1(a). Detailed information can be found elsewhere [10]. The X-ray diffraction reciprocal space mapping (XRD-RSM) was also conducted to analyze the crystallographic quality and strain state of the GeSn epi, as seen in Fig. 1(b). Each (Si)GeSn epi layer is correspondingly marked in the RMS figure, indicating excellent epitaxy quality.

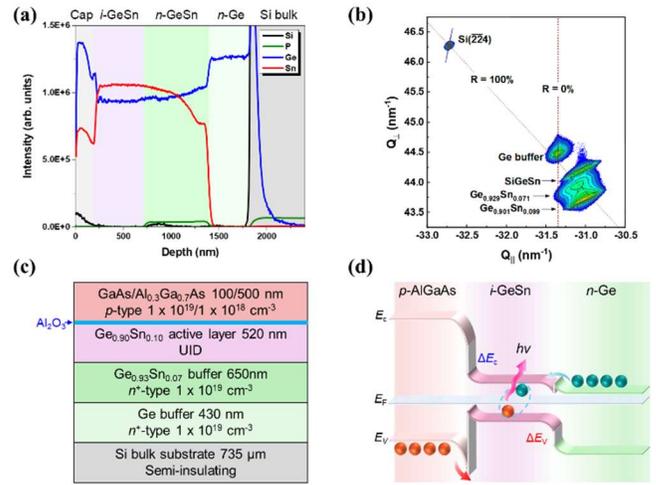

Fig. 1. (a) SIMS profile showing the distribution of elements within the structure, and (b) XRD-RSM of the as-grown GeSn substrate illustrating the crystallographic quality and strain state. (c) Diagram of the grafted AlGaAs/GeSn heterojunction, depicting the layer structure and interfaces. The 220 nm cap layer in the as-grown epi structure was removed using chemical mechanical polishing and was replaced with a GaAs/AlGaAs NM. (d) Energy band diagram of the AlGaAs/GeSn/Ge heterojunction at thermal equilibrium, highlighting the large conduction and valence band offsets.

As mentioned earlier, epitaxially grown GeSn heterojunctions present limited junction properties, particularly for DHS laser applications. Constrained by lattice matching requirements for CVD growth, typically only a few tens of meV in the conduction band offset ($\Delta E_C$) could be available. In the as-grown epi structure (Figs. 1(a) and 1(b)), the $\Delta E_C$ between epitaxial $Si_{0.03}Ge_{0.89}Sn_{0.08}$ cap and $Ge_{0.90}Sn_{0.10}$ active layers is ~0.114 eV [10], and the refractive index difference ($\Delta n$) is ~0.15 [23]. By replacing the original $Si_{0.03}Ge_{0.89}Sn_{0.08}$ cap with a foreign $Al_{0.3}Ga_{0.7}As$ layer (referred to as AlGaAs hereafter), as schematically shown in Fig. 1(c), the grafted AlGaAs/GeSn heterojunction presents more favorable electronic and optical properties resulted from the much larger band offsets and refractive indices than the as-grown epi structure.

At the AlGaAs/GeSn interface, a sub-nanometer thin layer of $Al_2O_3$ was introduced using atomic layer deposition (ALD). This amorphous oxide buffer layer serves multiple purposes: it passivates dangling bonds, reduces defects and interface traps, and effectively bonds to both material surfaces regardless of their lattice mismatch, without affecting charge carrier transport at the interface through quantum tunneling mechanisms [21]. Benefiting from this ultrathin ALD-$Al_2O_3$ interlayer, high-quality heterojunctions with ideal band alignment can be realized [16,24].

Fig. 1(d) illustrates the energy band diagram of the AlGaAs/GeSn/Ge heterojunction. The grafting of AlGaAs as a cap cladding layer in DHS laser design provides two main advantages. First, a larger conduction band offset $\Delta E_C$ between AlGaAs and GeSn can be expected. From our recent characterizations via X-ray photoelectron spectroscopy, the conduction band offset between

AlGaAs and GeSn is large as 1.19 eV [25], offering a higher electron barrier in the DHS structure and increasing the carrier recombination rate in the intrinsic active region. This electronic effect is also evidenced by the increased rectification ratio of a grafted AlGaAs/GeSn heterojunction diode (Fig. S1). Second, the increased refractive index difference $\Delta n$ (i.e., 1.01 versus 0.15) results in better optical confinement, further enhancing the light-matter interaction in the active layer (Fig. S2).

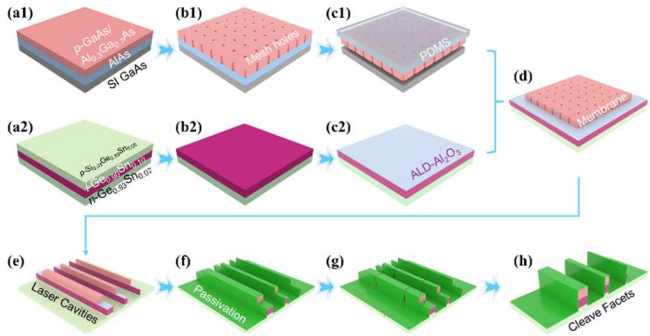

Fig. 2. Fabrication process of grafted AlGaAs/GeSn DHS laser. (a1) to (c1) Synthesis of GaAs/AlGaAs nanomembrane. (a2) to (c2) Preparation of GeSn substrate. (d) Formation of AlGaAs/GeSn/Ge heterostructure. (e) Creation of isolated laser cavities with varying device footprints. (f) Surface passivation of laser cavities via ALD-$Al_2O_3$. (g) Lapping and cleaving of fabricated lasers. (h) Finalized ridge lasers with a cavity length of ~1.1 mm.

**Laser fabrication**. The surface roughness of the as-grown GeSn epi (Fig. 2(a2)) was measured to be 14.52 nm using a Profilm3D® optical profiler (see Fig. S3(a)). To ensure a smooth interface and facilitate a high yield grafting process, chemical mechanical polishing (CMP) was performed, reducing the roughness to 1.92 nm, as measured by the same profiler (see Fig. S3(b)). This CMP process also removed the cap $p$-$Ge_{0.95}Sn_{0.05}$ layer (Fig. 2(b2)). Subsequently, an ALD tool was used to deposit a ~0.5-nm (5 cycles) thin layer of $Al_2O_3$ onto the GeSn substrate at a temperature of 200 °C (Fig. 2(c2)).

The transfer-printing of the GaAs/AlGaAs NM to the ALD-$Al_2O_3$ coated GeSn substrate begins with the separation of the NM from its bulk wafer (Fig. 2(a1)). This is achieved by patterning a grid of holes (9 × 9 $\mu m^2$) onto the top surface. A dry etch process is then used to etch through the exposed GaAs surface down to the AlAs layer (Fig. 2(b1)). The high aluminum content sacrificial layer is subsequently removed using a wet etch process with a 1:250 mixture of 49% HF and deionized (DI) water for 90 minutes. The detached NM is then picked up from the donor substrate using a polydimethylsiloxane (PDMS) polymer stamp and quickly rinsed in an alkaline solution of developer MIF321 to eliminate potential residues (Fig. 2(c1)) [26]. After cleaning the NM backsides, they are transfer-printed onto the host ALD-$Al_2O_3$-coated GeSn substrates. The samples are then thermally annealed using a rapid thermal annealer (RTA) at 200 °C for 5 minutes to ensure a strong chemical bond between the grafted NM and GeSn (Fig. 2(d)).

Following the grafting of the GaAs/AlGaAs NM, ridge structures were patterned onto the membrane surface, between the mesh hole patterns, using an optical contact lithography tool. The ridges were formed by etching through the GaAs/AlGaAs NM and intrinsic GeSn layers using separate ICP dry etch processes (Fig. 2(e)). To suppress surface recombination of the formed optical cavities, an ~8-nm thick (80 cycles) layer of $Al_2O_3$ was deposited using the same ALD system previously used for the interface oxide (Fig. 2(f)). To facilitate optical characterization of the fabricated ridge lasers, the samples were lapped down from backside to approximately 70 $\mu m$ and cleaved to a cavity length of about 1.1 mm (Figs. 2(g) and 2(h)).

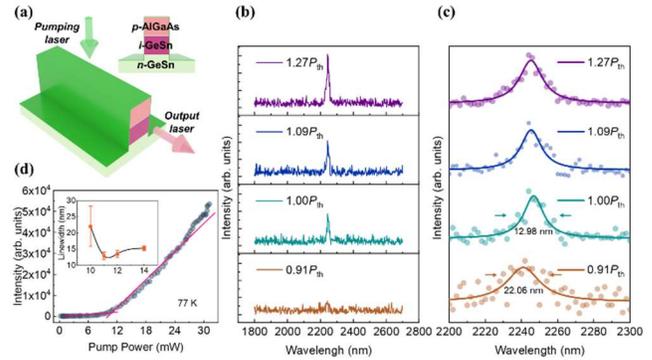

Fig. 3. Optical characterization of the grafted AlGaAs/GeSn/Ge DHS laser with a lateral size of 30 $\mu m$, operating at 77 K. (a) Schematic illustration of the optical pumping setup for the 30-$\mu m$ ridge cavity laser. (b) Stimulated laser spectra at various pumping states. (c) Lorentzian curve fitting of collected data below and above the lasing threshold, showing a clear linewidth narrowing. (d) L-L curve and linewidth of the lasing peak at ~2244 nm. The error bar in the inset represents the fitting uncertainty of the Lorentz model. The extracted lasing threshold at 77 K is ~11.06 mW, or 921.59 kW/$cm^2$.

**Laser characteristics.** As schematically represented in Fig. 3(a), the optical pumping characterization of the grafted laser was performed using a 1064 nm pulsed laser source focused into a spot size of ~20 μm in width and 3 mm in length, with 10 kHz repetition rate and 2 ns pulse width. Emission from the cavity facet was collected by a PbS detector, which has a detection range of 1.0 to 2.8 $\mu m$. The lasing spectra at various pumping powers from the grafted AlGaAs/GeSn laser, with a ridge cavity size of 30 $\mu m$, were measured at 77 K and are shown in Fig. 3(b). The magnified lasing peaks near ~2244 nm under different pumping conditions is presented in Fig. 3(c) and fitted using a Lorentzian function. The linewidth of the peak is estimated to be ~22.06 nm just below the threshold, while the linewidth of the spectra at the threshold shows a clear narrowing to ~12.98 nm. The collected intensity (L-L) curve in Fig. 3(d) exhibits a soft turn-on behavior, indicating a high spontaneous emission coupling efficiency [27]. Additionally, a low lasing threshold of ~11.06 mW (921.59 kW/$cm^2$) is extracted from the kink-shaped L-L curve. The inset of Fig. 3(d) summarizes the linewidth deduced from the Lorentzian fitting of lasing peaks at pumping states below, at, and above the threshold.

**Temperature-dependent characteristics.** To assess the thermal stability of our grafted AlGaAs/GeSn lasers, temperature-dependent photoluminescence (PL) spectra were collected, as shown in Fig. 4. Figure 4(a) presents the lasing spectra from 77 to 130 K at a pumping power of approximately 1.2 times the threshold ($P_{th}$) from the same 30-$\mu m$ cavity. As seen in Fig. 4(b), the lasing threshold increases with rising temperature from 77 to 130 K, which can be attributed to enhanced nonradiative recombination. To extract the characteristic temperature ($T_0$), the thresholds were fitted with an exponential function $P_{th} \propto Exp(T/T_0)$, resulting in $T_0 \approx$ 37.40 K. Meanwhile, Fig. 4(c) shows that the lasing wavelength red shifts with increasing temperature, demonstrating a differential shift of approximately 0.64 nm/K from a linear fit. This shift is likely related to the bandgap reduction in the GeSn active layer [11].

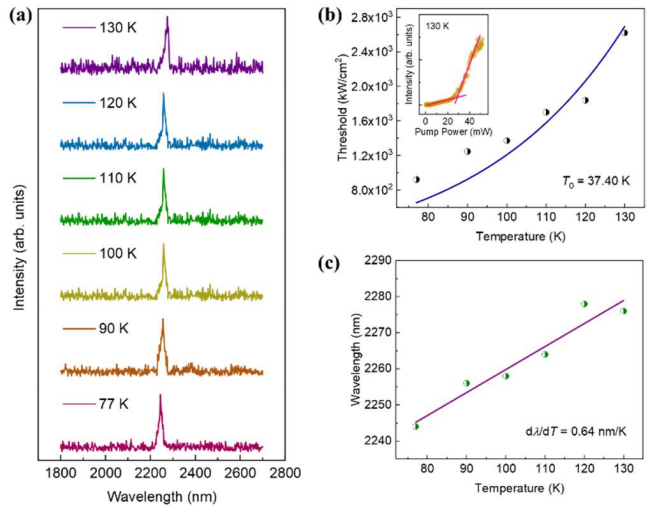

Fig. 4. Temperature-dependent characteristics of the grafted AlGaAs/GeSn/Ge DHS laser with a cavity size of 30 μm. (a) Normalized lasing spectra at various operation temperatures from 77 K to 130 K under a pumping power of ~1.2 $P_{th}$. (b) The relationship between lasing thresholds and temperatures. The inset shows the L-L curve at 130 K, indicating a threshold of 31.42 mW, or 921.59 kW/cm². (c) The variation of the lasing wavelength as a function of temperature.

**Lasing from a different cavity footprint.** In addition to the 30-μm cavity, we characterized another grafted laser on the same chip with a smaller ridge cavity size of 5 μm. The L-L curves and PL spectra at different temperatures are presented in Fig. 5. As expected, the collected lasing intensity in Fig. 5(a) demonstrates a decreasing trend with increasing temperature, and the lasing thresholds increase due to enhanced nonradiative recombination. The lasing behavior of this 5-μm cavity is observable up to 110 K, as shown in Fig. 5(b). The slightly reduced maximum operating temperature of the 5-μm cavity is likely due to the lessened gain medium volume in the active region. Additionally, as seen in Fig. 5(c), the linewidth of the lasing peaks exhibits a widening trend, indicating a deteriorating cavity quality factor at higher temperatures, as the quality factor Q is proportional to $\lambda/\Delta\lambda$, where $\lambda$ is the peak wavelength and $\Delta\lambda$ is the linewidth obtained from Lorentzian fitting.

**Conclusion.** In this study, we demonstrated the feasibility of using semiconductor grafting to develop group IV GeSn lasers. By grafting an AlGaAs nanomembrane as an electron blocking layer and an optical cladding layer, we achieved enhanced charge carrier and optical field confinement in GeSn edge-emitting lasers. The grafted lasers exhibited a lasing threshold of approximately 11.06 mW (921.59 kW/cm²) at 77 K and operated up to 130 K. These results highlight the potential of the grafting technique for advancing the development of electrically driven GeSn lasers capable of room-temperature operation.

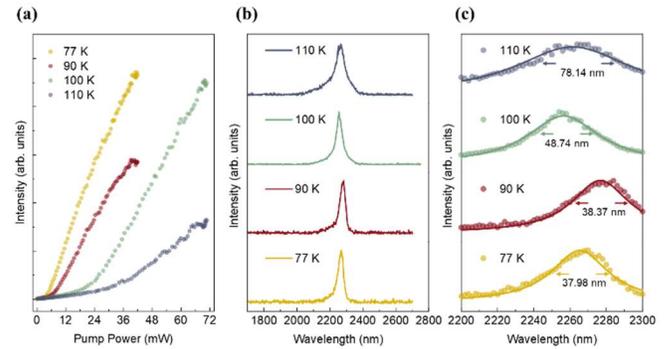

Fig. 5. Lasing characteristics from a different cavity with device footprint of 5 μm. (a) L-L curves of the laser at various operating temperatures. (b) Normalized lasing spectra recorded at temperatures ranging from 77 K to 110 K. (c) The variation of the lasing linewidth at different temperatures, fitted using a Lorentzian function.

**Funding.** The work was supported by Air Force Office of Scientific Research under Grant No. FA9550-19-1-0102. This work is supported in part by the National Science Foundation under Award No. 2235443.

**Disclosures.** The authors declare no conflicts of interest.

**Data availability**. Data underlying the results presented in this paper are not publicly available at this time but may be obtained from the authors upon reasonable request.

**Supplemental document.** See Supplement 1 for supporting content.

# GRAFTED ALGAAS/GESN OPTICAL PUMPING LASER OPERATING UP TO 130 K: SUPPLEMENTAL DOCUMENT

JIE ZHOU,[1,†] DANIEL VINCENT,[1,†] SUDIP ACHARYA,[2,3,†] SOLOMON OJO,[2,3,†] ALIREZA ABRAND,[4] YANG LIU,[1] JIARUI GONG,[1] DONG LIU,[1] SAMUEL HAESSLY,[1] JIANPING SHEN,[1] SHINING XU,[1] YIRAN LI,[1] YI LU,[1] HRYHORII STANCHU,[2] LUKE MAWST,[1] BRUCE CLAFLIN,[5] PARSIAN K. MOHSENI,[4,*] ZHENQIANG MA[1,*] AND SHUI-QING YU[2,*]

[1]Department of Electrical and Computer Engineering, University of Wisconsin - Madison, Madison, 53706, USA
[2]Department of Electrical Engineering, University of Arkansas, Fayetteville, Arkansas 72701, USA
[3]Microelectronics-Photonics Program, University of Arkansas, Fayetteville, Arkansas 72701, USA
[4]Department of Electrical and Microelectronic Engineering, Rochester Institute of Technology, Rochester, New York 14623, USA
[5]Air Force Research Laboratory, Wright-Patterson AFB, OH 45433, USA
[†]The authors contributed equally to this letter
*mazq@engr.wisc.edu
*syu@uark.edu
*pkmeen@rit.edu

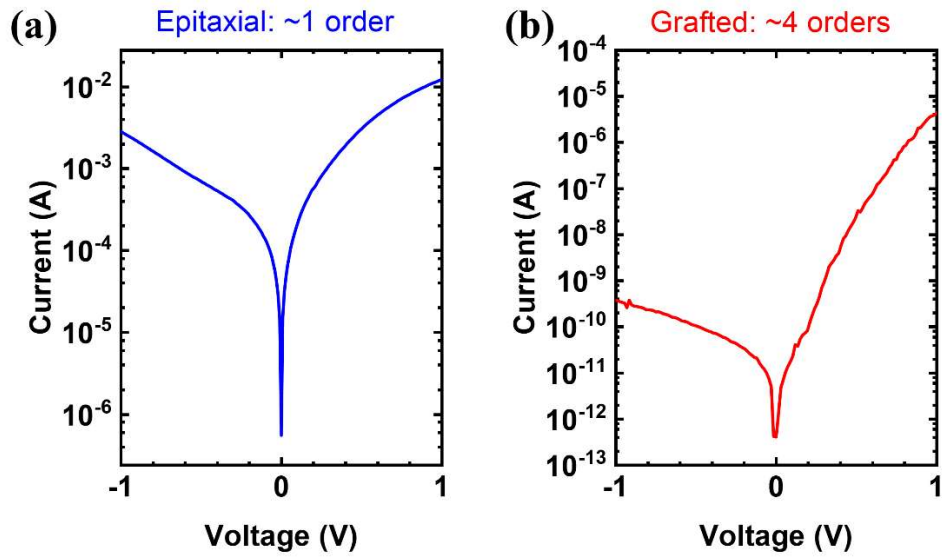

Fig. S1. Comparison of the rectifying behaviors of diodes fabricated from (a) as-grown epitaxial GeSn wafer, and (b) grafted AlGaAs/GeSn heterojunction. The rectification ratio increases from ~1 order of magnitude for the epitaxial reference diode to ~4 order of magnitude for the grafted diode, signifying the effect of an elevated band offset by grafting an AlGaAs layer.

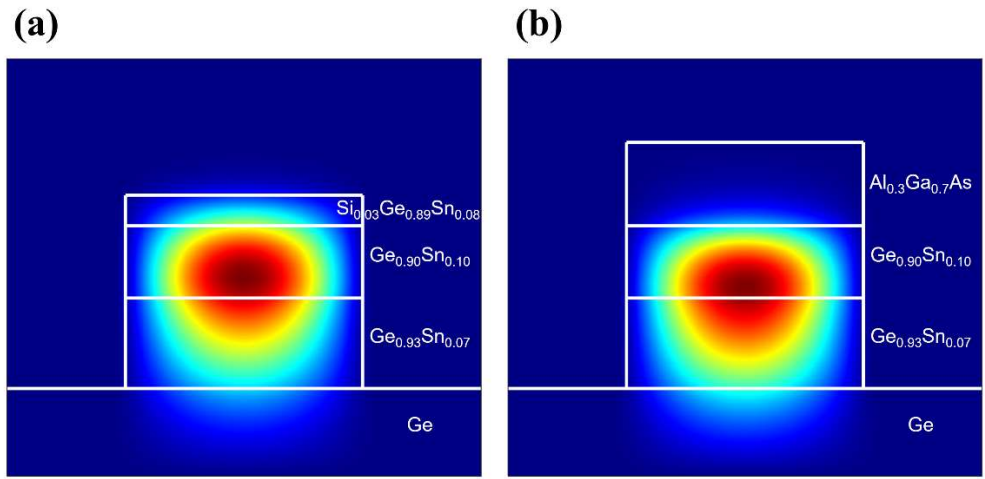

Fig. S2. FDTD simulation of optical field distributions (fundamental TE mode) within two different types of ridge cavities, including (a) as-grown $Si_{0.03}Ge_{0.89}Sn_{0.08}/Ge_{0.90}Sn_{0.10}/Ge_{0.93}Sn_{0.07}$, and (b) grafted $Al_{0.3}Ga_{0.7}As/Ge_{0.90}Sn_{0.10}/Ge_{0.93}Sn_{0.07}$ heterostructures. The larger difference of refractive index between AlGaAs and GeSn allows for a tighter light confinement and thus better light-matter interaction.

(a)
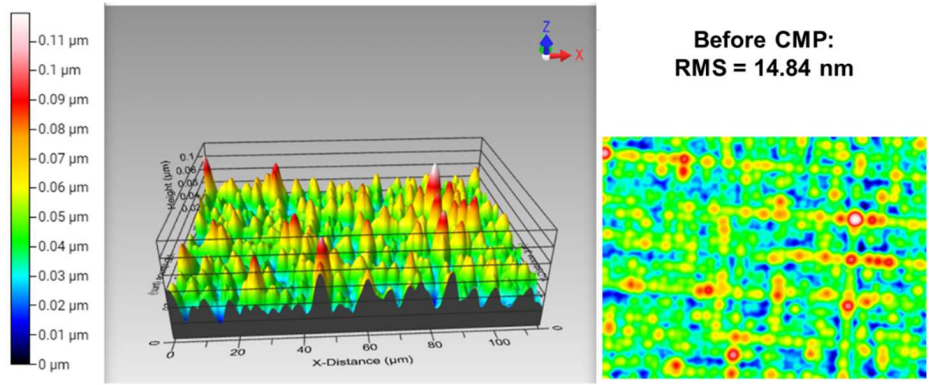

(b)
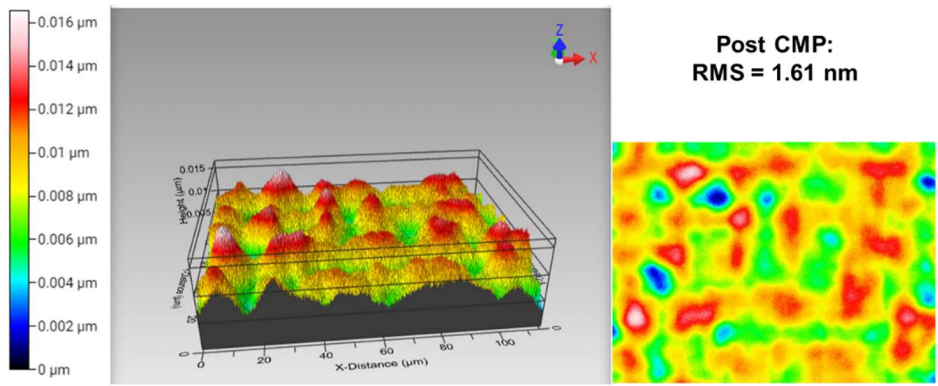

Fig. S3. Comparison of the surface morphology (a) before and (b) after chemical mechanical polishing (CMP), as captured using a Profilm3D® profilometer. The root mean square (RMS) roughness was reduced from 14.84 nm to 1.61 nm post CMP, ensuring that a high-yield membrane transfer-printing and smooth interface can be achieved.

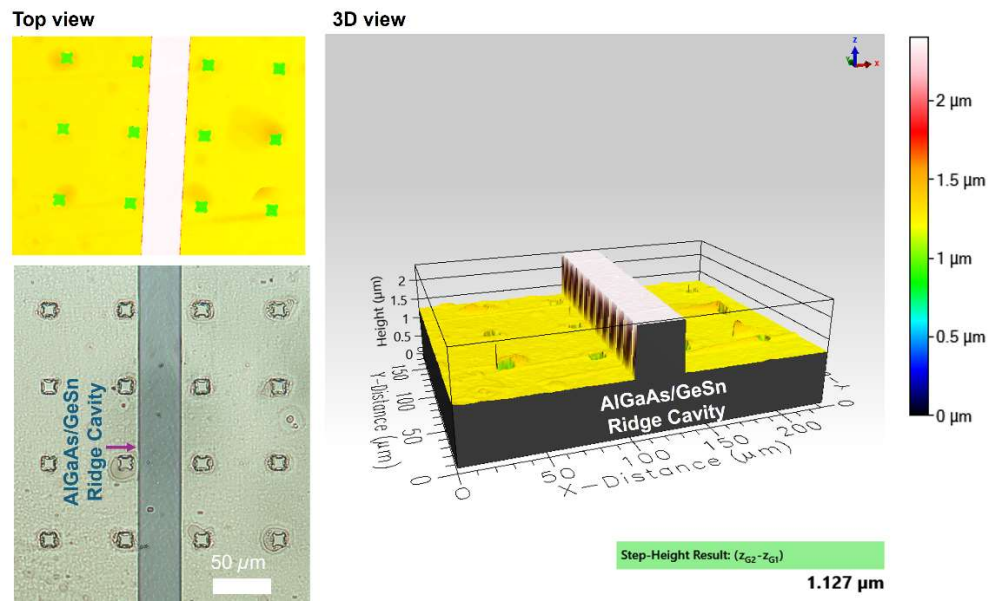

Fig. S4. Microscopic view and optical profiling of the fabricated grafted $Al_{0.3}Ga_{0.7}As/$ $Ge_{0.90}Sn_{0.10}/Ge_{0.93}Sn_{0.07}$ ridge cavity. The residual hole patterns on the GeSn substrate originate from the AlGaAs dry etching process for cavity formation.